\documentclass[aps,pra,twocolumn,superscriptaddress,showpacs,floatfix]{revtex4-1}
\usepackage{amsmath}
\usepackage{graphicx}
\usepackage{dcolumn}
\usepackage{bm}
\usepackage{amssymb}
\usepackage{mathrsfs}
\usepackage[colorlinks,
            linkcolor=blue,
            anchorcolor=blue,
            citecolor=blue,
            urlcolor=blue
            ]{hyperref}
\usepackage{xcolor}	 
\definecolor{myred}{RGB}{255,66,56}

\newcommand{\be}{\begin {equation}} 
\newcommand{\ee}{\end{equation}}
\newcommand{\mk}{\mathbf{k}}

\begin{document}

\title{Transition from nodal loop to nodal chain phase in a periodically modulated optical lattice}

\author{Chuanhao Yin}
\affiliation{Beijing National Laboratory for Condensed Matter Physics, Institute of Physics, Chinese Academy of Sciences, Beijing 100190, China}
\affiliation{School of Physical Sciences, University of Chinese Academy of Sciences, Beijing 100049, China}
\author{Linhu Li}
\affiliation{Beijing Computational Science Research Center, Beijing 100089, China}
\affiliation{CeFEMA, Instituto Superior
T\'ecnico, Universidade de Lisboa, Av. Rovisco Pais, 1049-001 Lisboa,  Portugal}
\author{Shu Chen}
\thanks{schen@iphy.ac.cn}
\affiliation{Beijing National Laboratory for Condensed Matter Physics, Institute of Physics, Chinese Academy of Sciences, Beijing 100190, China}
\affiliation{School of Physical Sciences, University of Chinese Academy of Sciences, Beijing 100049, China}
\affiliation{Collaborative Innovation Center of Quantum Matter, Beijing, China}

\begin{abstract}
We propose to study the transition from a nodal loop to nodal chain phase in a tunable two-dimensional $\pi$-flux lattice with periodical modulation potential. The Hamiltonian describes a periodically modulated optical lattice system under artificial magnetic fluxes and the tunable modulation phase factor provides additionally an artificial dimension of external parameter space. We demonstrate that this lattice system is able to describe a semimetal with either nodal loop or nodal chain Fermi surface in the extended three-dimensional Brillouin zone. By changing the strength of modulation potential $V$, we realize the transformation between the nodal-loop and nodal-chain semimetal.
\end{abstract}

\maketitle

\section{introduction}
During the past few years,  topological semimetals have attracted intensive studies in the field of condensed matter physics \cite{2007-Murakami-p356,2011-Wan-p205101,2011-Burkov-p127205,2011-Burkov-p235126,2015-Lv-p724,2015-Xu-p613,2015-Xu-p748,2015-Yang-p728,2011-Yang-p75129,2011-Xu-p186806,2012-Halasz-p35103,2012-Young-p140405,2015-Huang-p7373},
which opens a new window for exploring topological phases beyond the topological insulators and superconductors \cite{2010-Hasan-p3045,2011-Qi-p1057}.
Depending on the geometrical structure of the touching surface of conduction and valence bands, a topological semimetal may be a Weyl semimetal  or a nodal line semimetal. While a Weyl semimetal has discrete band-crossing points in the Brillouin zone, the band-crossing points of a nodal line semimetal generally form closed loops or rings \cite{2015-Fang-p81201,2016-Fang-p117106}, due to the periodicity of the Brillouin zone.
Several theoretical calculated materials have been proposed to realize nodal line semimetals, which have separated nodal-loop Fermi surfaces and host 'drumhead' surface states \cite{2015-Weng-p45108}. 
More recently, theoretical calculations unveil the existence of the  nodal chain semimetals in the material of iridum tetrafluoride (IrF$_4$) \cite{2016-Bzdusek-p75},  in AlB$_2$-type TiB$_2$ and ZrB$_2$ \cite{WuQS-2017}
and also W-C-type HfC \cite{2017-Yu-p36401}, 
with the nodal chain composed of several connected nodal rings. Experimental observation of nodal chains in a metallic-mesh photonic crystal has been reported very recently \cite{2017-Lu}. \par

Besides the progress for exploring topological semimetals  in real materials, significant advances have been made in some
artificial systems, such as photonic and acoustic crystals \cite{2013-Lu-p294,2015-Lu-p622,2015-Xiao-p920,2016-Chen-p13038}
and cold atomic systems \cite{2015-Dubfmmodeheckclseciek-p225301,2011-Lan-p165115,2012-Jiang-p33640,2012-Delplace-p67004,2015-Zhang-p13632,2017-Zhang-p43619,2016-He-p13606,2015-Xu-p265304,2016-Xu-p53619, 2016-Zhang-p43617,2016-Xu-p63606}. Schemes of realizing Weyl \cite{2015-Dubfmmodeheckclseciek-p225301,2011-Lan-p165115,2012-Jiang-p33640,2012-Delplace-p67004,2015-Zhang-p13632,2017-Zhang-p43619,2016-He-p13606,2015-Xu-p265304,2016-Xu-p53619}
and nodal-ring semimetals \cite{2016-Zhang-p43617,2016-Xu-p63606} 
in three-dimensional (3D) optical lattices with laser-assisted tunneling have
been theoretically proposed.
So far, the cold atomic systems provide a powerful platform with unparalleled controllability
towards studying topological states of matter. Some celebrated topological models, e.g., the Harper-Hofstadter
model \cite{1976-Hofstadter-p2239,1955-Harper-p874} 
and Haldane model \cite{1988-Haldane-p2015}, 
have been already realized \cite{2013-Miyake-p185302,2013-Aidelsburger-p185301,2015-Aidelsburger-p162,2014-Jotzu-p237} 
experimentally  in optical lattices.
In comparison with traditional condensed matter systems, the good tunability of cold atomic systems also provides feasible schemes for the simulation and detection of topological states with an artificial dimension in a lower dimensional optical lattice. A typical example system is the one-dimensional (1D) superlattice system, where the phase factor in the periodical or quasiperidocal modulation potential plays a role of artificial dimension and thus the 1D superlattice system is effectively connected to the two-dimensional (2D) Hall system \cite{2012-Lang-p220401}. 
If the phase factor is time-dependently tunable, it is possible to realize the topological charge pump in the 1D superlattice system \cite{2013-Wang-p26802,2017-Xu-p13606}, 
which has been experimentally observed recently \cite{2016-Lohse-p350,2016-Nakajima-p296}.

In this work, we  propose to study the nodal-line semimetal in a 2D optical lattice with periodical modulation potential. By changing the strength of the modulation potential, the transformation between nodal-line semimetal and nodal-chain semimetal occurs. The paper is organized as follows: In Sec.~\ref{model}, we introduce the periodically modulated model in the 2D optical lattice with a tunable parameter $\delta$. We calculate the energy dispersion and discuss the shape of Fermi surface under different value of modulation potential strength, which indicates a transition from a nodal-loop to nodal-chain semimetal. Density of state is also calculated to illustrate the transition.
Finally, we give a brief summary in Sec.~\ref{Summary}.
\par

\section{Model and results} \label{model}

\begin{figure}[tb]
\centering
\includegraphics[width=0.45\textwidth]{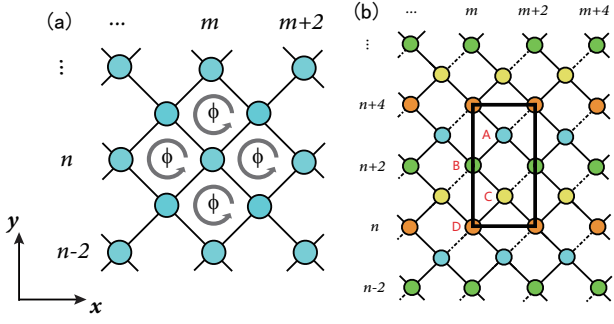}
\caption{A schematic diagram of optical lattice. (a) Two dimension optical lattice with a phase $\phi$ acquired in each cell. (b) Periodic modulated optical lattice in the $y$ direction with four kinds sublattice (A, B, C, D for four kinds of color: blue, green, yellow, orange).
$m,n$ represent coordinates in the $x$ and $y$ direction, respectively. Solid black lines represent the hopping processes with phase $0$ and dash black lines represent the processes with phase $\pi$. }\label{fig1}
\end{figure}

We first consider a 2D square lattice model with a spatially varying hopping phase, as described by the Hamiltonian
\be
H_t = -t\sum_{\langle m,n \rangle}(e^{-i\phi_{m,n}} \hat a^\dagger_{m+1,n+1} \hat a_{m,n} + \hat a^\dagger_{m-1,n+1} \hat a_{m,n} + h.c.)
\ee
with $ \hat a$ ($ \hat a^\dagger$) the annihilation (creation) operator and $t$ the hopping amplitude.
The summation of $\langle m,n \rangle$ runs over all lattice points.
A flux of $\phi$ can be acquired in each square by choosing $\phi_{m,n}=n\phi$, which simulates a perpendicular uniform magnetic field, as shown in Fig.\ref{fig1}(a). This model has been experimentally realized in optical lattice with two pairs of  counter-propagating laser beams in $x+y$ and $x-y$ directions \cite{2013-Miyake-p185302,2013-Aidelsburger-p185301}. 
Starting from this model, we introduce additionally a third pair of laser beams in the $y$ direction, which addresses a periodically modulated on-site potential along the $y$ axis and can be described by
\be
H_V = V\sum_{(m,n)}  \cos(2\pi\alpha n + \delta) \hat n_{m,n},
\ee
where $\hat n_{m,n} =  \hat a_{m,n}^\dagger  \hat a_{m,n}$, $V$ is the strength of the potential, $\alpha$ is a rational number that determines the period of the potential, and $\delta$ is an arbitrary phase. As $\delta$ can continuously vary from $0$ to $2 \pi$,  it provides an artificial dimension to simulate a higher-dimensional system via dimension extension \cite{2012-Lang-p220401,2012-Kraus-p106402,2015-Zhang-p13632}. 

The unit cell of this model is determined by $\phi$ and $\alpha$ together, and we shall focus on the specific case with $\phi = \pi$ and $\alpha = 1/4$ in this work. In such a case, the model describes a periodically modulated $\pi$-flux square lattice with four sublattices as shown in Fig.\ref{fig1}(b). For simplicity, here we take $m$ and $n$ as even numbers and thus the phase acquired by hopping distributes as displayed in Fig.\ref{fig1}(b) with the flux in each square being $\pi$. Now we rewrite the Hamiltonian as
\be
H = H_t + H_V,
\ee
with
\begin{align}
H_t =& -t\sum_{unit~cell}\big[ e^{-i\pi} \hat a_{m+2,n+4}^{D\dagger} \hat a_{m+1,n+3}^A+ \hat a_{m,n+2}^{B\dagger} \hat a_{m+1,n+3}^A  \nonumber \\
& + \hat a_{m,n+4}^{D\dagger} \hat a_{m+1,n+3}^A + \hat a_{m+2,n+2}^{B\dagger} \hat a_{m+1,n+3}^A \nonumber \\
&+e^{-i\pi} \hat a_{m+2,n+2}^{B\dagger} \hat a_{m+1,n+1}^C+ \hat a_{m,n}^{D\dagger} \hat a_{m+1,n+1}^C \nonumber \\
&+ \hat a_{m,n+2}^{B\dagger} \hat a_{m+1,n+1}^C + \hat a_{m+2,n}^{D\dagger} \hat a_{m+1,n+1}^C \big] + h.c., \nonumber
\end{align}
and
\begin{align}
H_V &= \sum_{unit~cell} [V_1 \hat n^A_{m+1,n+3}+ V_2 \hat n^B_{m,n+2} \nonumber \\
 &+ V_3 \hat n^C_{m+1,n+1}+ V_4 \hat n^D_{m,n}], \nonumber
\end{align}
where $V_j=V\cos{(j\pi/2+\delta)}$ and $\hat n^\beta_{m,n}= \hat a_{m,n}^{\beta \dagger} \hat a_{m,n}^\beta$, where $\beta = A,B,C,D$. The summation $\sum_{unit~cell}$ takes every unit cell as shown by the black rectangle in Fig.\ref{fig1}(b), i.e.
Without loss of generality, we set the hopping amplitude $t=1$ hereafter. \par
Next, we could use Fourier transform $\hat a_{m,n}^\beta = \frac{1}{\sqrt{L_x L_y}} \sum_{k_x,k_y} e^{i(k_x ma+k_y na)} \hat a_{k_x,k_y}^\beta $, with $a$ the lattice constant. $k_x$ and $k_y$ are wave vectors defined in the Brillouin region of $L_x\times L_y$ square lattice. For simplicity, we take $a=1$ hereafter. The Hamiltonian can be written in momentum space as
\begin{align}
H(\mk) =& H_t(\mk) + H_V(\mk) \nonumber ,\\
H_t(\mk)  =&-t\sum_{\mk} \left[ M  \hat a_{\mk}^{D\dagger} \hat a_{\mk}^A + N  \hat a_{\mk}^{B\dagger} \hat a_{\mk}^A + M  \hat a_{\mk}^{B\dagger} \hat a_{\mk}^C \right. \nonumber \\
 & \left.+ N  \hat a_{\mk}^{D\dagger} \hat a_{\mk}^C \right]+h.c.  ,\nonumber \\
H_V(\mk)  = & \sum_{\mk}\left[V_1  \hat a_{\mk}^{A\dagger} \hat a_{\mk}^A + V_2  \hat a_{\mk}^{B\dagger} \hat a_{\mk}^B
+ V_3  \hat a_{\mk}^{C\dagger} \hat a_{\mk}^C + V_4  \hat a_{\mk}^{D\dagger} \hat a_{\mk}^D \right], \nonumber
\end{align}%
with $M = e^{i k_y}2\cos{k_x}$ and $N = -e^{-i k_y}2\sin{k_x}$. For convenience, we rewrite the Hamiltonian as
\be\label{H_k_matrix}
H(\mk)= \hat \psi_{\mk}^\dagger h(\mk) \hat \psi_{\mk}, 
\ee
with
$
\hat \psi_{\mk}=\left( \hat a_{\mk}^A , \hat a_{\mk}^B , \hat a_{\mk}^C , \hat a_{\mk}^D \right) ^T, \nonumber
$
and
\be
h({\mathbf{k}}) = \left(
\begin{array}{cccc}
-V\sin \delta  & M^* & 0 & N^* \\
M & -V\cos \delta  & N & 0 \\
0 & N^* & V\sin \delta  & M^* \\
N & 0 & M & V\cos \delta
\end{array}%
\right).  \nonumber
\ee
By diagonalizing Eq.(\ref{H_k_matrix}), the energy dispersion can be achieved as
\begin{widetext}
\begin{equation}\label{E4}
E(\mk,\delta)=\pm \sqrt{4
+\frac{V^{2}}{2}\pm \sqrt{ 16 \sin^2 2k_x \cos^2 2k_y + 4V^2 + 4V^2 \sin 2\delta \cos 2k_x + \frac{V^{4}}{4} \cos^2 2\delta}}.
\end{equation}
\end{widetext}

\begin{figure}[tb]
\centering
\includegraphics[width=0.5\textwidth]{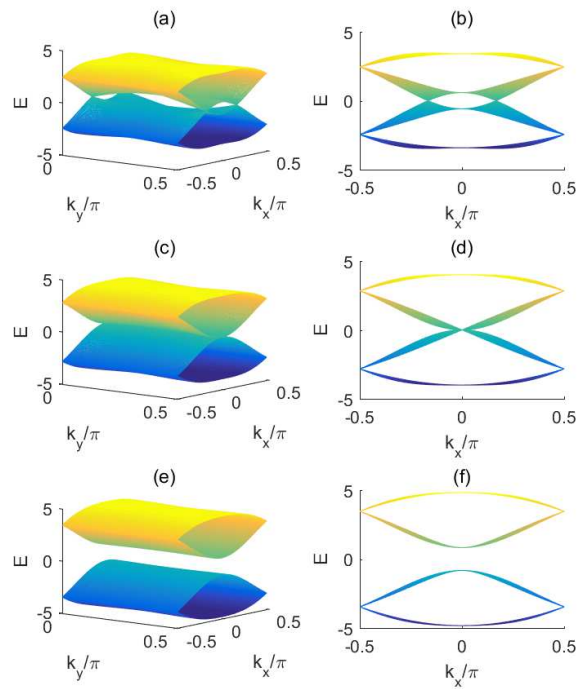}
\caption{The spectrum of the system for different value of $V$ at $\delta = \pi/4$. From top to bottom, the values of $V$ are (a,b) $2$, (c,d) $\sqrt{8}$ and (e,f) $4$, respectively. Figure (b), (d), (f) is the left view of left figure (a), (c), (e), correspondingly.}\label{fig2}
\end{figure}

From the expression of dispersion, we see that the energy dispersion of Eq.(\ref{E4}) is symmetric about zero energy. Hence the system at half-filling falls into a metallic phase if $E(\mk,\delta)=0$ is satisfied in certain region of the Brillouin zone, or an insulating phase otherwise. The zero energy condition yields
\be \label{relation}
\left\{
\begin{array}{c}
\sin 2k_x \sin 2k_y=0, \\
V^{2}\sin 2\delta - 8\cos 2k_x = 0.
\end{array}%
\right.
\ee
It is clear that the above condition is fulfilled when $k_y=0,\pi$ and $\cos{2 k_x}=-\frac{1}{8} V^2\sin{2\delta}$, which give rise to two 2D Dirac points. Furthermore, one observes that Eqs.(\ref{relation}) is also fulfilled when $k_x=0,\pi$  and $ V^2\sin{2\delta}=8$.
To see it clearly, we consider a concrete system with fixed $\delta=\pi/4$ and demonstrate the energy spectrum of the system for various $V$ in Fig,\ref{fig2}.
Here we only display the first Brillouin zone with $k_x\in[-\pi/2,\pi/2]$ and $k_y\in[0,\pi/2]$, as the energy dispersion fulfills $E(k_x,k_y,\delta) = E(k_x+\pi,k_y,\delta) = E(k_x,k_y+\frac{\pi}{2},\delta) $. The dispersion also fulfills $E(k_x,k_y,\delta) = E(k_x,k_y,\delta+\pi)$, which introduces a period of $\pi$ for the parameter $\delta$. \par

From Fig.\ref{fig2}, we see that the half-filled system with a fixed $\delta$ carries out a transition from a semimetal to an insulator phase when the strength $V$ is increased.
When $V<\sqrt{8/\sin{2\delta}}$, the system is a semimetal with two Dirac points. While varying $V$  moves positions of the Dirac points in the Brillouin zone,  they merge together at $k_x=0$ when $V=\sqrt{8/\sin{2\delta}}$. As shown in Fig. \ref{fig2}(c), the merging points form a nodal line $\mathbf{L}_1$, corresponding to the solution of Eqs.(\ref{relation}):  $k_x=0$ and $V=\sqrt{8/\sin{2\delta}}$, which is independent of the value of $k_y$. For $V>\sqrt{8/\sin{2\delta}}$, Eqs.(\ref{relation}) are no longer fulfilled, and a gap is opened. Varying $\delta$ will modify the critical value of $V$, which tends to infinity when $\delta=0$ or $\pi/2$.

\begin{figure*}[tb]
\centering
\includegraphics[width=1.0\textwidth]{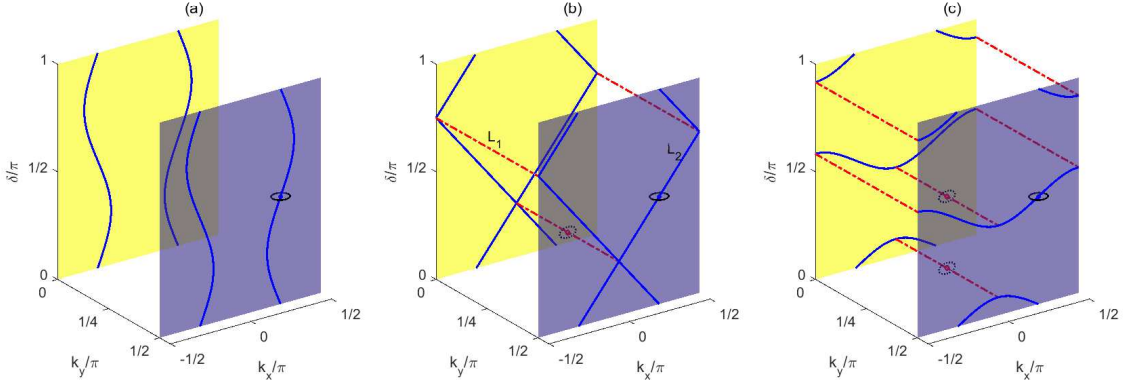}
\caption{Nodal lines in the extended-3D Brillouin zone with different potential strength, (a) $V=2$, (b) $V=\sqrt{8}$ and (c) $V=4$. The red  dash-dotted and blue solid lines are for the nodal lines of $\mathbf{L}_1$ and $\mathbf{L}_2$, respectively, and the black dash and solid circles are the integral path of $\gamma_n$ for corresponding nodal lines.} \label{fig3}
\end{figure*}

As $\delta$ can be continuously changed from $0$ to $2\pi$, it provides an additional parameter dimension with tunability.  Thus we can take $\delta$ as a quasi-momentum, and study the structure of nodal points in an effective 3D Brillouin zone of $(k_x,k_y,\delta)$. The separated Dirac points located at $k_y=0$ or $k_y=\pi/2$ will form some 1D lines $\mathbf{L}_2$ along the third direction of $\delta$, as shown by the blue solid lines in Fig.\ref{fig3}. Here we only show the region with $\delta\in[0,\pi]$ as the spectrum has a period of $\pi$ in $\delta$.  Since $\delta$ is now considered as a quasi-momentum, the quasi-3D model  has only one variable parameter $V$, while the gap closing condition is provided by Eqs.(\ref{relation}).

The change of the structure of nodal lines for various $V$ is shown in Fig.\ref{fig3}.
When $V<\sqrt{8}$, there only exists one set of nodal lines $\mathbf{L}_2$ in $k_x-\delta$ space at $k_y=0$ and $\pi/2$. Increasing $V$  modifies the shape of $\mathbf{L}_2$, and these lines touch together at $(k_x,\delta) = (0,\pi/4)$ and $(\pm\pi/2,3\pi/4)$  when $V=\sqrt{8}$, as shown in Fig.\ref{fig3}(b). Meanwhile, the other set of nodal lines $\mathbf{L}_1$ emerges and connects the nodal lines of $\mathbf{L}_2$. As shown in Fig.\ref{fig3}(b), the nodal lines of $\mathbf{L}_1$ and $\mathbf{L}_2$ are chained together to form a nodal chain.  When $V>\sqrt{8}$, the nodal chain is split into two separated nodal chains with each nodal chain composed of $\mathbf{L}_1$ and $\mathbf{L}_2$. As shown in Fig.\ref{fig3}(c), the nodal lines $\mathbf{L}_1$ always connect with $\mathbf{L}_2$ at places of $k_x=0$ and $k_x=\pm \pi/2$, and the lines of $\mathbf{L}_2$ go through the Brillouin zone along the $k_x$ direction, whereas they go through the Brillouin zone along the $\delta$  direction when $V<\sqrt{8}$. When $V$ tends to infinity, each separated nodal chain becomes flat as nodal lines $\mathbf{L}_1$ only locate on the same surface with $\delta =0~(\pi)$ or  $\delta=\pi/2$.
It is clear that a transition occurs at $V=\sqrt{8}$, corresponding to the structure of Fermi surface in the extended 3D Brillouin zone changing from nodal lines to nodal chains. We note that a negative $V$ gives similar results, as the potential strength enters the spectrum relation as $V^2$.\par
\par

\begin{figure}[tb]
\centering
\includegraphics[width=0.5\textwidth]{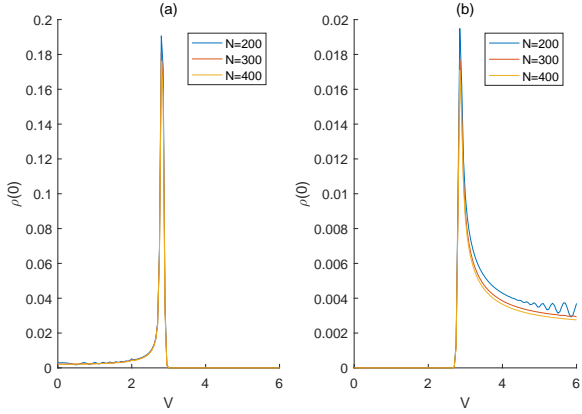}
\caption{(a) Density of state for the half-filled system with fixed $\delta=\pi/4$. (b) Density of state for the half-filled system with fixed $k_y = \pi/4$. The different color of lines in (a) and (b) represents different value of lattice size $N$.} \label{fig4}
\end{figure}
The topological properties of a nodal line can be characterized by a Berry phase of $\pi$ along a trajectory enclosing the line in the quasi-momentum space. Here we define this trajectory as a small circle in a certain plane with a phase angle $\theta$, which encloses the crossing point of the nodal line and the given plane \cite{2015-Fang-p81201}.  The Berry phase for the $n$th band is defined as
\be
\gamma_n = \oint  A_n(\theta) d\theta,
\ee
with the Berry connection $A_n(\theta)$ given by
\be
A_n(\theta) = -i \langle u_n(\theta)|\partial_\theta| u_n(\theta)\rangle
\ee
and $|u_n(\theta)\rangle$ representing the $n$th eigenstate of the model. In our model, the nodal lines $\mathbf{L}_1$ and $\mathbf{L}_2$ are the touching region of the $2$nd and $3$rd bands, hence we can characterize these lines solely by $\gamma_2$. For the nodal lines $\mathbf{L}_2$, we consider a small circle in the $k_x-k_y$ plane at $\delta=\pi/2$ (the black solid circles in Fig.\ref{fig3}), with the center of the circle located at the point of $(k_x,k_y)=(\pi/4,\pi/2)$ as $\mathbf{L}_2$ always goes through this point. Integrating along the chosen path $(\pi/4+A\cos(\theta), \pi/2+A\sin(\theta))$ with $A$ the radius of the circle, we achieve $\gamma_2=\pi$ for all three different $V$. On the other hand, the Berry phase along a trajectory enclosing a nodal line of $\mathbf{L}_1$ can be chosen on the $k_x-\delta$ plane, which is perpendicular to $\mathbf{L}_1$,
as displayed by the black dash line in Fig.\ref{fig3}(b) and (c). Numerical results give $\gamma_2=\pi$ also for each nodal line of $\mathbf{L}_1$ as long as $V>\sqrt{8}$. When $V=\sqrt{8}$, each pairs of nodal lines of $\mathbf{L}_1$ merge into one line, and the corresponding Berry phase is $0$ with modulo $2\pi$.

To see the change of Fermi surface more clearly, we also plot the density of state (DOS) versus $V$ for the half-filled system at some given parameters shown in Fig.~\ref{fig4}. The DOS is defined as
\begin{align}
\rho(E) =& \frac{1}{2N^2}\sum_{i=1}^{2N^2}\delta(E-E_i)\nonumber\\
=& \frac{1}{2N^2}\sum_{l=1}^{2}\sum_{i_1=-\frac{N}{2}}^{\frac{N}{2}-1}\sum_{i_2=-\frac{N}{2}}^{\frac{N}{2}-1}\delta(E-E_{l,i_1,i_2}).
\label{dos}
\end{align}
Here $E_{l,i_1,i_2}$ is the $l-th$ eigenstate with fixed $i_1$ and $i_2$. $i_1$ and $i_2$ represent different parameters in Fig.\ref{fig4}(a) and (b). In Fig.\ref{fig4}(a), we fix $\delta=\pi/4$ and take $k_x = \frac{\pi}{N}i_1$ and $k_y = \frac{\pi}{2N}i_2$. In Fig.\ref{fig4}(b), we fix $k_y=\pi/2$ and take $k_x = \frac{\pi}{N}i_1$ and $\delta = \frac{\pi}{N}i_2$. To numerically calculate $\delta(E-E_i)$, we make an approximation of the function of $\delta(x)$ by using a Gaussian function $\frac{1}{\sqrt{\pi\sigma^2}}exp(-\frac{x^2}{\sigma^2})$, 
which approaches the $\delta$-function exactly when $\sigma \rightarrow 0$.
For insulators, $\rho(0)$ must be zero. The appearance of nonzero $\rho(0)$ means the emergence of gap closing points. In Fig.\ref{fig4}(a), we plot the DOS at $E=0$ versus $V$ by fixing $\delta=\pi/4$. With the increase of the lattice size $N$, the curves become more and more smooth and $\rho(0)$ becomes zero when $V>\sqrt{8}$. A sharp peak appears at $V=\sqrt{8}$, which also gives a signature of transition from nodal line to nodal chain. Similarly, in Fig.\ref{fig4}(b) we plot $\rho(0)$ versus $V$ by fixing $k_y=\pi/2$. At $V=\sqrt{8}$, there also exists a sharp peak, corresponding to the transition point from nodal line to nodal chain.

\section{Summary} \label{Summary}
In summary, we have studied a periodically modulated 2D $\pi$-flux square lattice model with a tunable phase factor $\delta$ of the periodical modulation potential, which can be taken as a parameter of the extended spacing dimension. The model is realizable by engineering the Raman-assisted hopping of ultracold atoms in a 2D optical lattice with additional laser beams for the modulation in the $y$ direction. By analyzing the structure of Fermi surface, we find that the half-filled system can be either a node-line or node-chain semimetal and demonstrate a transition occurring at $V=\sqrt{8}$. When $V<\sqrt{8}$, the system is a nodal-loop semimetal, whereas it becomes a nodal-chain semimetal when $V>\sqrt{8}$.  The density of state at $E=0$ also gives signature of the transformation from the nodal-loop to nodal-chain semimetal.

\begin{acknowledgments}
The work is supported by the National Key Research and Development Program of China (2016YFA0300600), NSFC under Grants No. 11425419, No. 11374354 and No. 11174360, and the Strategic Priority Research Program (B) of the Chinese Academy of Sciences  (No. XDB07020000).
\end{acknowledgments}


\begin{references}
\bibitem{2007-Murakami-p356} S. Murakami, New Journal of Physics \textbf{9}, 356 (2007).
\bibitem{2011-Wan-p205101} X. Wan, A. M. Turner, A. Vishwanath, and S. Y. Savrasov, Phys. Rev. B \textbf{83}, 205101 (2011).
\bibitem{2011-Burkov-p127205} A. A. Burkov and L. Balents, Phys. Rev. Lett. \textbf{107}, 127205 (2011).
\bibitem{2011-Burkov-p235126} A. A. Burkov, M. D. Hook, and L. Balents, Phys. Rev. B \textbf{84}, 235126 (2011).
\bibitem{2015-Lv-p724} B. Q. Lv, N. Xu, H. M. Weng, J. Z. Ma, P. Richard, X. C. Huang, L. X. Zhao, G. F. Chen, C. E. Matt, F. Bisti, V. N. Strocov, J. Mesot, Z. Fang, X. Dai, T. Qian, M. Shi, and H. Ding, Nat Phys \textbf{11}, 724 (2015).
\bibitem{2015-Xu-p613} S.-Y. Xu, I. Belopolski, N. Alidoust, M. Neupane, G. Bian, C. Zhang, R. Sankar, G. Chang, Z. Yuan, C.-C. Lee, S.-M. Huang, H. Zheng, J. Ma, D. S. Sanchez, B. Wang, A. Bansil, F. Chou, P. P. Shibayev, H. Lin, S. Jia, and M. Z. Hasan, Science \textbf{349}, 613 (2015).
\bibitem{2015-Xu-p748} S.-Y. Xu, N. Alidoust, I. Belopolski, Z. Yuan, G. Bian, T.-R. Chang, H. Zheng, V. N. Strocov, D. S. Sanchez, G. Chang, C. Zhang, D. Mou, Y. Wu, L. Huang, C.-C. Lee, S.-M. Huang, B. Wang, A. Bansil, H.-T. Jeng, T. Neupert, A. Kaminski, H. Lin, S. Jia, and M. Zahid Hasan, Nat Phys \textbf{11}, 748 (2015).
\bibitem{2015-Yang-p728} L. X. Yang, Z. K. Liu, Y. Sun, H. Peng, H. F. Yang, T. Zhang, B. Zhou, Y. Zhang, Y. F. Guo, M. Rahn, D. Prabhakaran, Z. Hussain, S.-K. Mo, C. Felser, B. Yan, and Y. L. Chen, Nat Phys \textbf{11}, 728 (2015).
\bibitem{2011-Yang-p75129} K.-Y. Yang, Y.-M. Lu, and Y. Ran, Phys. Rev. B \textbf{84}, 075129 (2011).
\bibitem{2011-Xu-p186806} G. Xu, H. Weng, Z. Wang, X. Dai, and Z. Fang, Phys. Rev. Lett. \textbf{107}, 186806 (2011).
\bibitem{2012-Halasz-p35103} G. B. Hal\'asz and L. Balents, Phys. Rev. B \textbf{85}, 035103 (2012).
\bibitem{2012-Young-p140405} S. M. Young, S. Zaheer, J. C. Y. Teo, C. L. Kane, E. J. Mele, and A. M. Rappe, Phys. Rev. Lett. \textbf{108}, 140405 (2012).
\bibitem{2015-Huang-p7373} S.-M. Huang, S.-Y. Xu, I. Belopolski, C.-C. Lee, G. Chang, B. Wang, N. Alidoust, G. Bian, M. Neupane, C. Zhang, S. Jia, A. Bansil, H. Lin, and M. Z. Hasan, Nat. Commun. \textbf{6}, 7373 (2015).
\bibitem{2010-Hasan-p3045} M. Z. Hasan and C. L. Kane, Rev. Mod. Phys. \textbf{82}, 3045 (2010).
\bibitem{2011-Qi-p1057} X.-L. Qi and S.-C. Zhang, Rev. Mod. Phys. \textbf{83}, 1057 (2011).
\bibitem{2015-Fang-p81201} C. Fang, Y. Chen, H.-Y. Kee, and L. Fu, Phys. Rev. B \textbf{92}, 081201 (2015).
\bibitem{2016-Fang-p117106} C. Fang, H. Weng, X. Dai and Z. Fang, Chinese Physics B \textbf{25}, 117106 (2016).
\bibitem{2015-Weng-p45108} H. Weng, Y. Liang, Q. Xu, R. Yu, Z. Fang, X. Dai, and Y. Kawazoe, Phys. Rev. B \textbf{92}, 045108 (2015).
\bibitem{2016-Bzdusek-p75} T. Bzdu\v{s}ek, Q. Wu, A. R\"{u}egg, M. Sigrist, and A. A. Soluyanov, Nature \textbf{538}, 75 (2016).
\bibitem{WuQS-2017} X. Feng, C. Yue, Z. Song, Q. Wu and B. Wen, arXiv:1705.00511.
\bibitem{2017-Yu-p36401} R. Yu, Q. Wu, Z. Fang, and H. Weng, Phys. Rev. Lett. \textbf{119}, 036401 (2017).
\bibitem{2017-Lu} Q. Yan, R. Liu, Z. Yan, B. Liu, H. Chen, Z. Wang and L. Lu, arXiv:1706.05500.
\bibitem{2013-Lu-p294} L. Lu, L. Fu, J. D. Joannopoulos, and M. Soljacic, Nat Photon \textbf{7}, 294 (2013).
\bibitem{2015-Lu-p622} L. Lu, Z. Wang, D. Ye, L. Ran, L. Fu, J. D. Joannopoulos, and M. Solja{\v c}i{\'c}, Science \textbf{349}, 622 (2015).
\bibitem{2015-Xiao-p920} M. Xiao, W.-J. Chen, W.-Y. He, and C. T. Chan, Nat Phys \textbf{11}, 920 (2015).
\bibitem{2016-Chen-p13038} W.-J. Chen, M. Xiao, and C. T. Chan, Nat. Commun. 7, 13038 (2016).
\bibitem{2015-Dubfmmodeheckclseciek-p225301} T. Dub\ifmmode \check{c}\else \v{c}\fi{}ek, C. J. Kennedy, L. Lu, W. Ketterle, M. Solja\ifmmode \check{c}\else \v{c}\fi{}i\ifmmode \acute{c}\else \'{c}\fi{}, and H. Buljan, Phys. Rev. Lett. \textbf{114}, 225301 (2015).
\bibitem{2011-Lan-p165115} Z. Lan, N. Goldman, A. Bermudez, W. Lu, and P. {\"Ohberg}, Phys. Rev. B \textbf{84}, 165115 (2011).
\bibitem{2012-Jiang-p33640} J.-H. Jiang, Phys. Rev. A \textbf{85}, 033640 (2012).
\bibitem{2012-Delplace-p67004} P. Delplace, J. Li, and D. Carpentier, EPL (Europhysics Letters) \textbf{97}, 67004 (2012).
\bibitem{2015-Zhang-p13632} D.-W. Zhang, S.-L. Zhu, and Z. D. Wang, Phys. Rev. A \textbf{92}, 013632 (2015).
\bibitem{2017-Zhang-p43619} D.-W. Zhang, R.-B. Liu, and S.-L. Zhu, Phys. Rev. A \textbf{95}, 043619 (2017).
\bibitem{2016-He-p13606} W.-Y. He, S. Zhang, and K. T. Law, Phys. Rev. A \textbf{94}, 013606 (2016).
\bibitem{2015-Xu-p265304} Y. Xu, F. Zhang, and C. Zhang, Phys. Rev. Lett. \textbf{115}, 265304 (2015).
\bibitem{2016-Xu-p53619} Y. Xu and L.-M. Duan, Phys. Rev. A \textbf{94}, 053619 (2016).
\bibitem{2016-Zhang-p43617} D.-W. Zhang, Y. X. Zhao, R.-B. Liu, Z.-Y. Xue, S.-L. Zhu, and Z. D. Wang, Phys. Rev. A \textbf{93}, 043617 (2016).
\bibitem{2016-Xu-p63606} Y. Xu and C. Zhang, Phys. Rev. A \textbf{93}, 063606 (2016).
\bibitem{1976-Hofstadter-p2239} D. R. Hofstadter, Phys. Rev. B \textbf{14}, 2239 (1976).
\bibitem{1955-Harper-p874} P. G. Harper, Proceedings of the Physical Society. Section A \textbf{68}, 874 (1955).
\bibitem{1988-Haldane-p2015} F. D. M. Haldane, Phys. Rev. Lett. \textbf{61}, 2015 (1988).
\bibitem{2013-Miyake-p185302} H. Miyake, G. A. Siviloglou, C. J. Kennedy, W. C. Burton, and W. Ketterle, Phys. Rev. Lett. \textbf{111}, 185302 (2013).
\bibitem{2013-Aidelsburger-p185301} M. Aidelsburger, M. Atala, M. Lohse, J. T. Barreiro, B. Paredes, and I. Bloch, Phys. Rev. Lett. \textbf{111}, 185301 (2013).
\bibitem{2015-Aidelsburger-p162} M. Aidelsburger, M. Lohse, C. Schweizer, M. Atala, J. T. Barreiro, S. Nascimbene, N. R. Cooper, I. Bloch, and N. Goldman, Nat Phys \textbf{11}, 162 (2015).
\bibitem{2014-Jotzu-p237} G. Jotzu, M. Messer, R. Desbuquois, M. Lebrat, T. Uehlinger, D. Greif, and T. Esslinger, Nature \textbf{515}, 237 (2014).
\bibitem{2012-Lang-p220401} L.-J. Lang, X. Cai, and S. Chen, Phys. Rev. Lett. \textbf{108}, 220401 (2012).
\bibitem{2013-Wang-p26802} L. Wang, M. Troyer, and X. Dai, Phys. Rev. Lett. \textbf{111}, 026802 (2013).
\bibitem{2017-Xu-p13606} Z. Xu, Y. Zhang, and S. Chen, Phys. Rev. A \textbf{96}, 013606 (2017).
\bibitem{2016-Lohse-p350} M. Lohse, C. Schweizer, O. Zilberberg, M. Aidelsburger, and I. Bloch, Nat Phys \textbf{12}, 350 (2016).
\bibitem{2016-Nakajima-p296} S. Nakajima, T. Tomita, S. Taie, T. Ichinose, H. Ozawa, L. Wang, M. Troyer, and Y. Takahashi, Nat Phys \textbf{12}, 296 (2016).
\bibitem{2012-Kraus-p106402} Y. E. Kraus, Y. Lahini, Z. Ringel, M. Verbin, and O. Zilberberg, Phys. Rev. Lett. \textbf{109}, 106402 (2012).
\end{references}

\end{document}